\documentclass{ws-ijmpd}
\usepackage[super,compress]{cite}
\begin{document}

\markboth{C. Pittori}
{AGILE results on relativistic outflows above 100 MeV}

\catchline{}{}{}{}{}
%
%%%%%%%%%%%%%%%%%%%%%%%%%%%%%%%%%%%%%%%%%%%%%%%%%%%%%%%%%%%%%%%%%%%%

\title{AGILE RESULTS on RELATIVISTIC OUTFLOWS above 100 MeV}

\author{CARLOTTA PITTORI\footnote{on behalf of the AGILE Collaboration.}}

\address{INAF-OAR and ASI Space Science Data Center \\
Via del Politecnico snc, 00133 Rome, Italy \\
carlotta.pittori@ssdc.asi.it}

%%\author{SECOND AUTHOR}

%%\address{Group, Laboratory, Address\\
%%City, State ZIP/Zone, Country\\
%%second\_author@group.com}

\maketitle

\begin{history}
\received{Day Month Year}
\revised{Day Month Year}
\end{history}

\begin{abstract}
We give an overview of the AGILE $\gamma$-ray satellite highlights.
AGILE is an Italian Space Agency (ASI) mission devoted to 
observations in the 30 MeV - 50 GeV  $\gamma$-ray energy range, 
with simultaneous X-ray imaging in the 18-60 keV band.
Launched in April 2007, the AGILE satellite has completed its 
tenth year of operations in orbit, and it is substantially contributing 
to improve our knowledge of the high-energy sky.
Emission from cosmic sources at energies above 100 MeV 
is intrinsically non-thermal, and the study of the wide 
variety of observed Galactic and extragalactic $\gamma$-ray sources 
provides a unique opportunity to test theories of particle 
acceleration and radiation processes in extreme conditions.
\end{abstract}


\ccode{PACS numbers: 95.85.Pw, 95.55.Ka, 98.70.Rz, 98.38.Fs, 98.58.Fd.}

%\tableofcontents

\section{Introduction}	
\begin{figure}[pt]
%\centerline{\psfig{file=pittori_c_fig01.eps,width=10cm}}
\centerline{\psfig{file=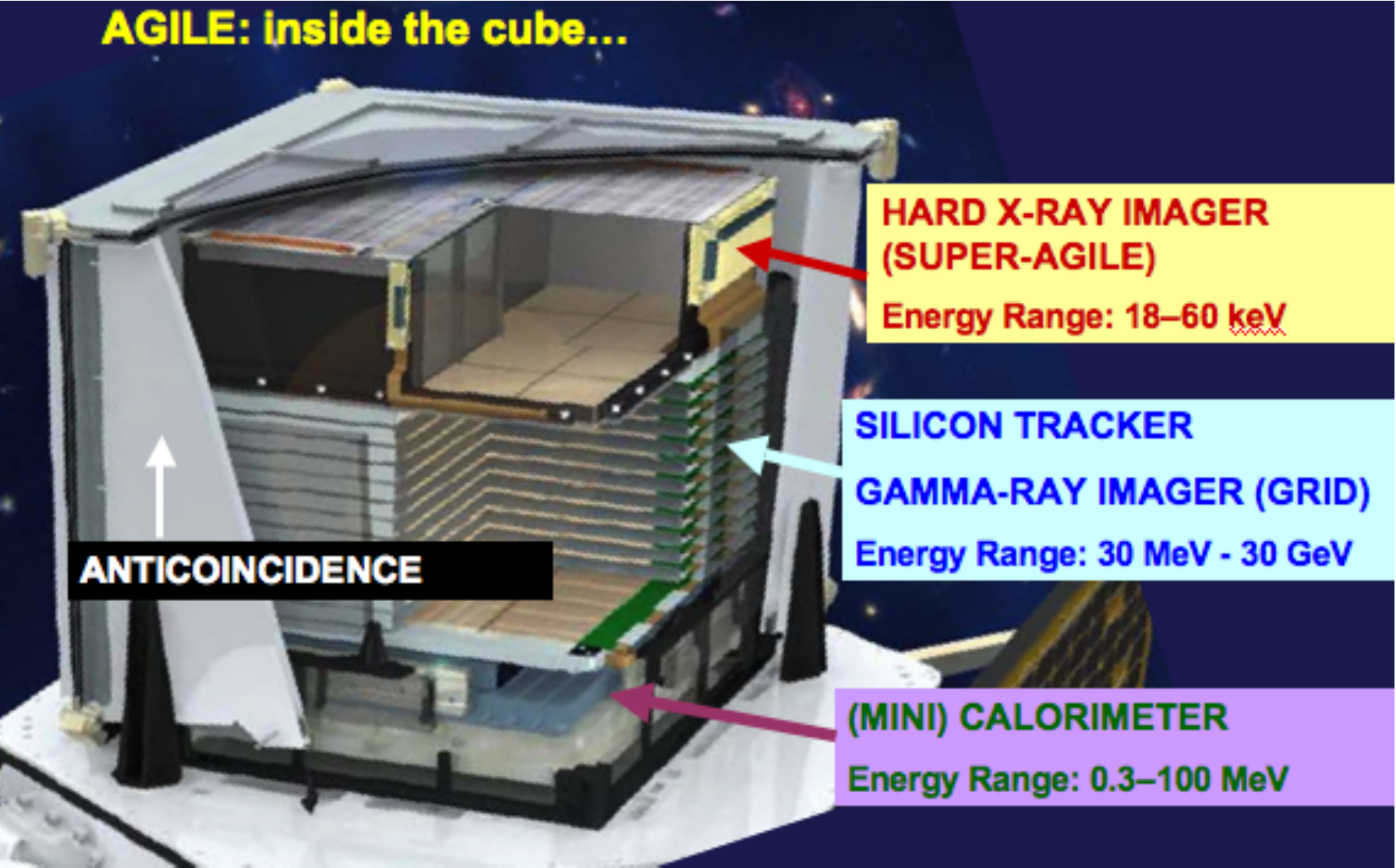,width=10cm}}
%\vspace*{8pt}
\caption{The AGILE payload.
\label{instrument}
}
\end{figure}

AGILE (Astro-rivelatore Gamma a Immagini LEggero) \cite{Tavani2009a}    %(Tavani et al., 2009) 
is a $\gamma$-ray astrophysics mission of the Italian Space Agency (ASI), 
with INAF, INFN and CIFS participation.
The satellite, launched on April 23, 2007, has completed its tenth year of operations, and continues its mission with high efficiency.
The AGILE scientific payload consists of three instruments with 
independent detection capability (see Fig.~\ref{instrument}): 
the Gamma Ray Imager Detector (GRID) sensitive in the 
energy range 30 MeV--30 GeV, a Mini-Calorimeter (MCAL) 
sensitive in the energy range 350 keV to 100 MeV
that works both as a slave of the GRID and as an autonomous 
detector for transient events, and a hard X-ray imager on top
(Super-AGILE), sensitive in the energy range 18--60 keV.

%%%%%%%%%%%%%%%%%%%%%%%%%%%%%%%%%%%%
High-energy astrophysics studies the non-thermal emission in the Universe, mainly
coming from violent astrophysical environments near compact objects, such as neutron stars
and super-massive or stellar-size black holes.
This branch of astrophysics has enjoyed rapid development in the past decades, 
and recent  important  results  and  progress  
were achieved by the space-based $\gamma$-ray AGILE \cite{Tavani2009a} 
and Fermi \cite{Atwood2009} (NASA) satellites observations.
Recent and unexpected discoveries of 
intense $\gamma$-ray transients at energies above 100 MeV on short timescales
($<$ minutes, days) in different astrophysical systems, both Galactic 
and extragalactic, challenge models of particle acceleration.
% Galactic compact objects, Quasars and relativistic jets, GRBs. \\
High-energy astrophysics new lessons:
%role of the magnetic field, super-acceleration, plasma instabilities.
the observation of brief and very bright flares of energetic $\gamma$-rays suggests that pairs are accelerated 
to PeV energies on short timescales. Such rapid intra-day accelerations cannot be driven by shocks,
and they may highlight the role of the magnetic field, the importance of relativistic 
magnetic reconnection and plasma instabilities in astrophysical sources.

Furthermore the compact objects emitting broad-band non-thermal electromagnetic radiation
are also believed to be emitters of other multi-messenger signals,
such as cosmic rays, neutrinos, and gravitational waves.
%%%%%%%%%%%%%%%%%%%%%%%%%%%%%%%%%%%%%%%%%%%%%%%%%%%%%%%%%%

\section{AGILE Science Highlights}
\label{sec:highlights}

We present here a selection of the main AGILE science highlights 
after ten years of operations, and some recent updates related to 
the search of electomagnetic counterparts of gravitational waves (GW).

In particular we reassume in Subsect. \ref{sec:cyg}, \ref{sec:w44} and \ref{sec:crab}
the three main discoveries for which AGILE will be remembered.
In Subsect. \ref{sec:other} we present other important scientific results 
for which we think that AGILE {\it should} be remembered.

\subsection {AGILE detections of microquasars in the Cygnus region}

\begin{figure}[pt]
\centerline{\psfig{file=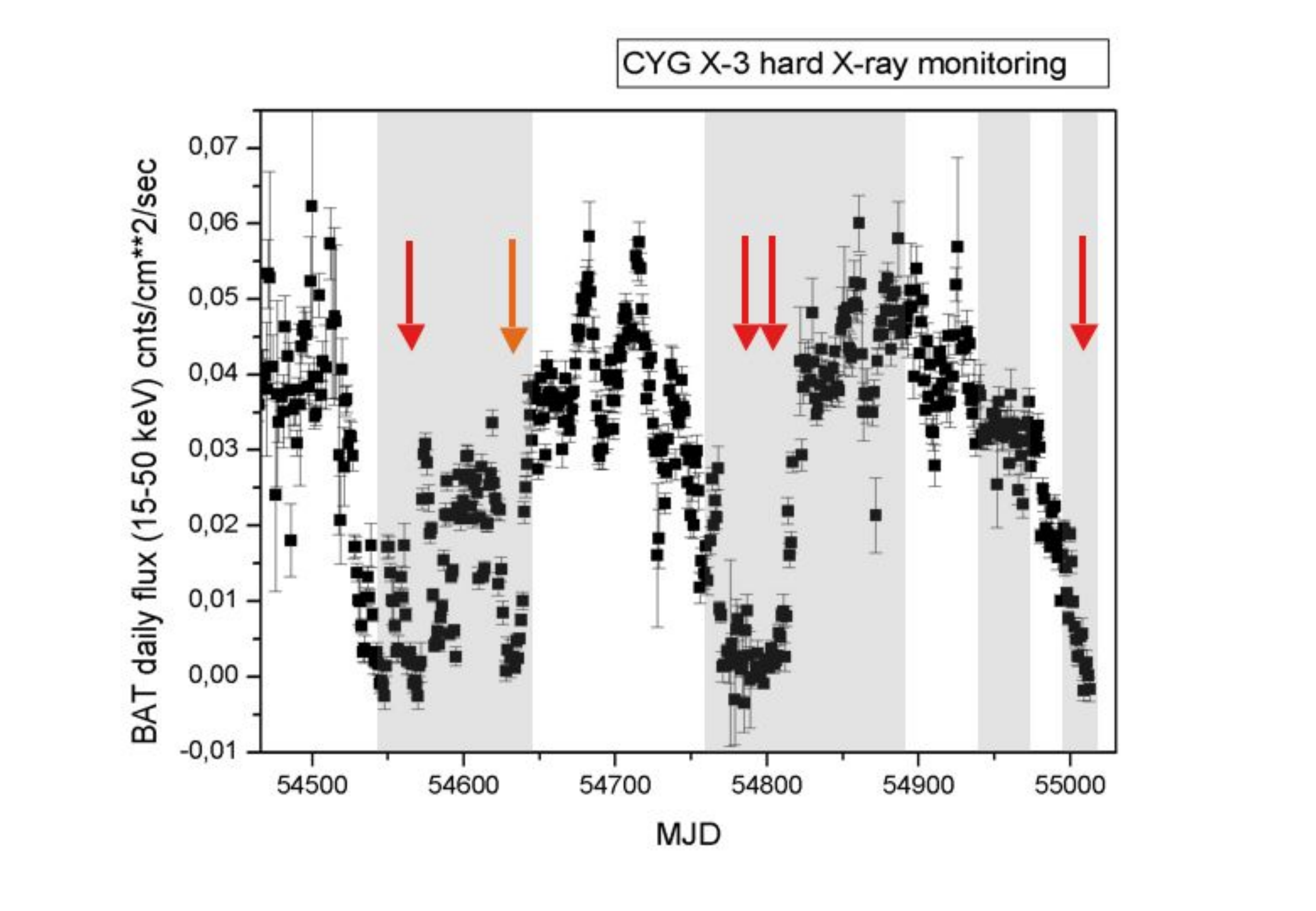,width=14cm}}
%\vspace*{8pt}
\caption{Hard-X-ray flux from Cygnus X-3 as monitored by the Burst
Alert Telescope (BAT) on board NASA’s Swift spacecraft, between 1
January 2008 and 30 June 2009. The red arrows mark the dates of
major gamma-ray flares of Cygnus X-3 as detected by the AGILE
instrument \cite{Tavani2009b}.
\label{CygX-3}
}
\end{figure}

\label{sec:cyg}
\begin{itemlist}
\item {\bf Cygnus X-3}:
The AGILE discovery of  transient $\gamma$-ray emission from Cygnus X-3
in 2008 associated with a specific spectral state preceding
a major radio jet ejection opened a new window of investigation
of microquasars.
AGILE detected for the first time several $\gamma$-ray flares above 100 MeV from
Cygnus X-3 microquasar and also a weak persistent emission \cite{Tavani2009b}. 
%(Tavani et al., 2009, Nature). 
Gamma-ray flares occur following a clear repetitive pattern,
either in coincidence with low hard-X-ray fluxes or during transitions
from low to high hard-X-ray fluxes, see Fig. \ref{CygX-3}, and usually appear 
before major radio flares.
This important AGILE discovery published on the high-impact
journal {\it Nature} has been subsequently confirmed  
on a {\it Science} paper by Fermi, which was also able to identify the 4.8 hours orbital period 
in $\gamma$-rays, securing unambigously the temporal signature of the binary system \cite{Abdo2009}.
% (Abdo et al., 2009, Science).
AGILE and Fermi were able to answer a long-lasting question:
Cygnus X-3 binary system is indeed able to accelerate particles 
up to relativistic energies and to emit $\gamma$-rays above 100 MeV.

\item {\bf Cygnus X-1}:
%In 2010, $\gamma$-
Gamma-ray flaring activity for a source positionally consistent with Cygnus X-1 
microquasar was reported twice by AGILE in 2010 \cite{Bulgarelli2010,Sabatini2010a,Sabatini2010b}.
%(Bulgarelli et al., 2010; Sabatini et al., 2010a and 2010b).
AGILE extensive monitoring of Cygnus X-1 in the 
energy range 100 MeV - 3 GeV during the period 2007
July - 2009 October confirmed the existence of a 
spectral cutoff between 1-100 MeV during the typical hard
X-ray spectral state of the source. However, even in this state,
Cygnus X-1 is capable of producing episodes of extreme particle 
acceleration on 1-day timescales,
and even shorter lived flares in the TeV range as detected by MAGIC in 2006 
{Albert2007}.
%(Albert et al., 2007).
Gamma-ray flares of Cygnus X-1 detected by AGILE above 100 MeV were not immediately
confimed by Fermi, and have been confirmed only after a later re-analisis of Fermi-LAT data 
\cite{Bodaghee2013}.
%(Bodaghee et al., 2013).
\end{itemlist}

\subsection{First evidence of cosmic-ray acceleration from AGILE observations of the SNR W44}
\label{sec:w44}

\begin{figure}[pt]
\vspace*{-48pt}
\centerline{\psfig{file=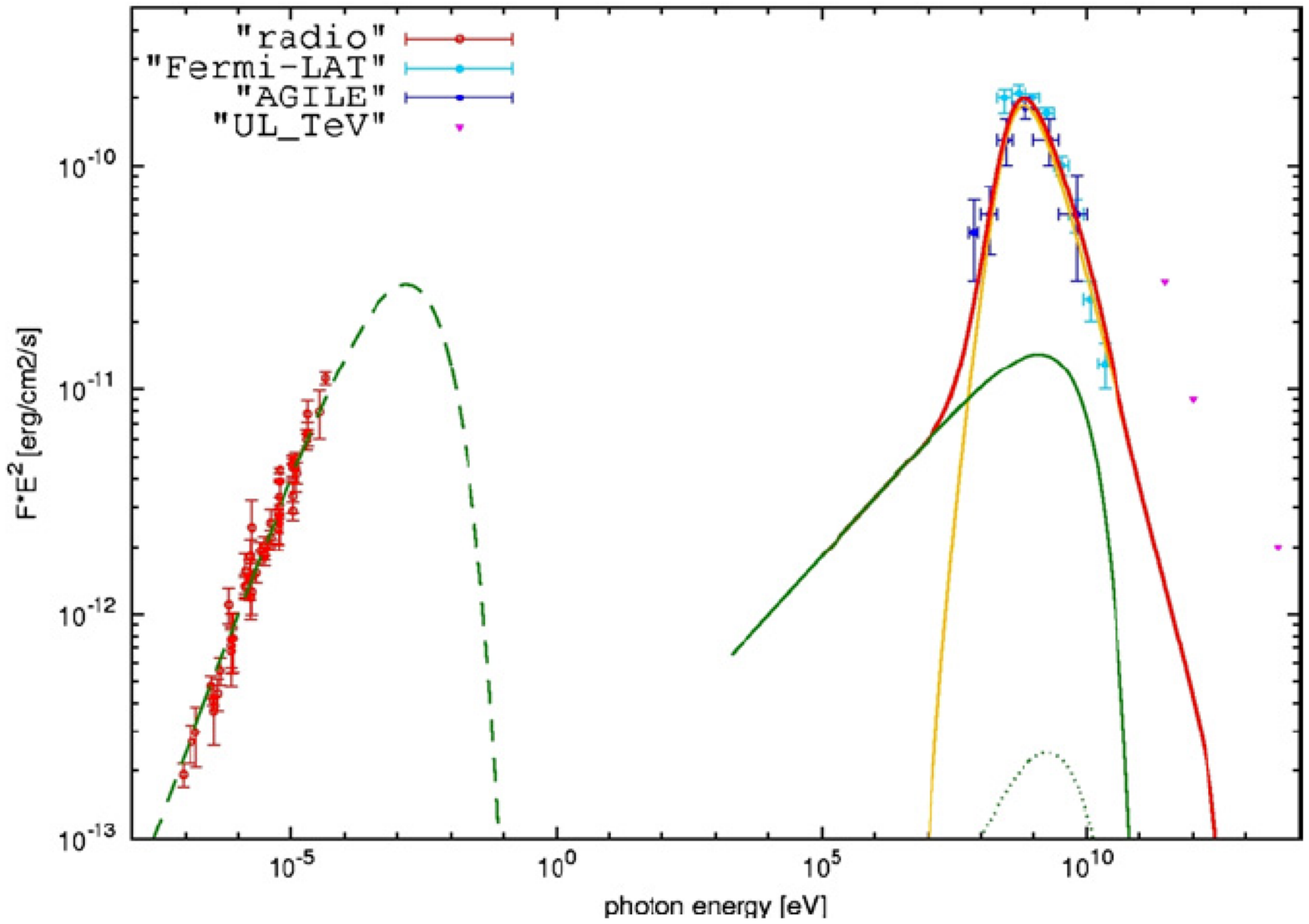,width=18cm}}
\vspace*{-48pt}
\caption{Hadronic modeling of the broadband spectral energy distribution of the SNR W44
with radio data in red and gamma-ray data in blue \cite{Giuliani2011}.
The yellow curve shows the neutral pion emission from the accelerated proton distribution.
\label{Giulianiw44}
}
\end{figure}

Understanding the origin of cosmic rays is one of the most important 
issues of high-energy astrophysics, and galactic Supernova Remnants (SNR) are considered to be 
an ideal laboratory to study Cosmic-Ray acceleration.
In the energy range just below 200 MeV,
hadronic and leptonic emission spectra have a well distinct behavior due to a steepening of the
hadronic spectrum due to the neutral pion emission, which is missing in the leptonic case.
The AGILE-GRID instrument reaches its optimal sensitivity 
just at energies between 50 MeV and a few GeV,
and it was the first to discover a clear indication of the so-called ``pion bump''
in the $\gamma$-ray 
emission from the supernova remnant W44 \cite{Giuliani2011}.
AGILE observations, combined with the observed multifrequency properties of the 
source, have been crucial to discriminate between theoretical models, and
can be unambiguously attributed to accelerated protons interacting 
with nearby dense gas, see Fig. \ref{Giulianiw44}.
This important AGILE result was reported in \refcite{Giuliani2011} %(Giuliani et al., 2011), 
and later confirmed by Fermi-LAT data in \refcite{Ackermann2013}.  %(Ackermann et al., 2013).
Before AGILE and Fermi, a direct identification of proton acceleration sites in our Galaxy 
was elusive. 

\subsection{Crab Nebula variability}
\label{sec:crab}

\begin{figure}[pt]
\vspace*{-48pt}
\centerline{\psfig{file=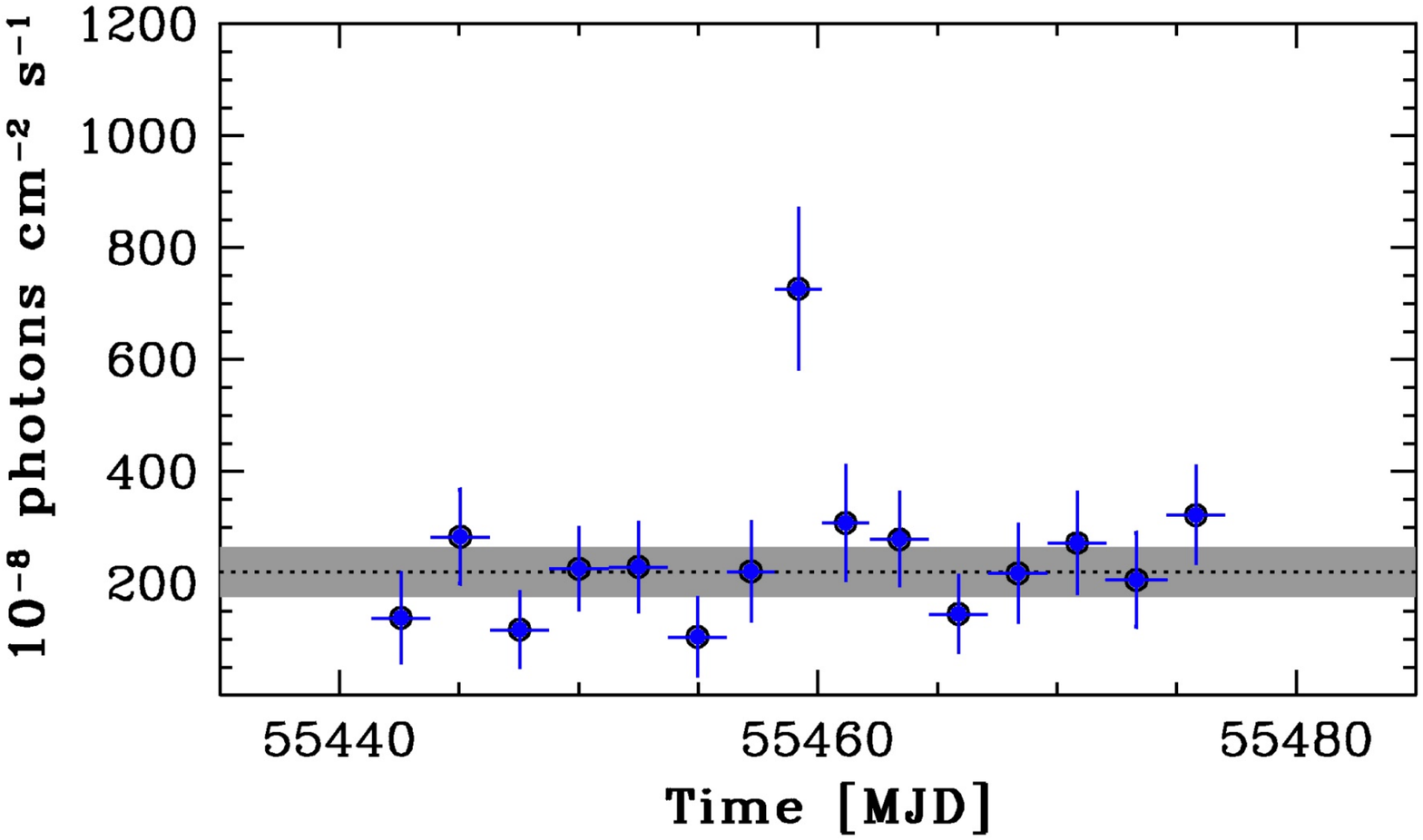,width=15cm}}
\vspace*{-48pt}
\caption{The Crab Nebula flare as observed by
AGILE at energies above 100 MeV in September 2010
\cite{Tavani2011}.
\label{Crab2010}
}
\end{figure}

The surprising discovery by AGILE of variable $\gamma$-ray emission
above 100 MeV from the Crab Nebula in Sept. 2010 \cite{Tavani2010,Tavani2011}
%(Tavani et al., 2010; Tavani et al., 2011) 
started a new era of investigation of the Crab system.
The 2012 Bruno Rossi International Prize has been awarded to the PI, 
Marco Tavani, and the AGILE team for this important
and unexpected discovery, which was also confirmed
one day later by Fermi \cite{Buehler2010,Abdo2011}.
%(Buehler et al., 2010; Abdo et al., 2011).
%

Crab was considered to be an almost ideal standard candle, a nearly
constant source (at a level of few percent) from optical to $\gamma$-ray energies, 
with possible long-term nebular
flux variations over a few-year timescale reported in the hard X-ray range.
On September 2010 AGILE detected a rapid 
$\gamma$-ray flare over a daily timescale, see Fig.\ref{Crab2010}, and thanks to its rapid alert system, 
made the first public announcement on September 22, 2010 \cite{Tavani2010}.
%M. Tavani et al., Astron. Telegram 2855 (2010).
This finding was confirmed the next day by the Fermi Observatory  \cite{Buehler2010}.
%R. Buehler et al., Astron. Telegram 2861 (2010).
AGILE had also previoulsy detected a giant flare from the Crab in October, 2007 
during the initial Science Verification Period of the satellite, and
in the First AGILE Catalog paper \cite{Pittori2009} %(Pittori et al., 2009)
it was reported that anomalous episodic flux values observed
from the Crab in 2007 were under investigation. 
We know now of several major $\gamma$-ray flares from the Crab
Nebula detected by the AGILE-GRID and Fermi-LAT, up to October 2016.

Gamma-ray data provide evidence for particle acceleration mechanisms 
in nebular shock regions more efficient than previously expected 
from theoretical models. We estimate a recurrence rate for
strong $\gamma$-ray flares of $\sim 1/$year.

\subsection{Other important AGILE scientific results}
\label{sec:other}
\begin{itemlist}
\item {\bf PWN Vela-X}:
The AGILE detection of $\gamma$-ray emission from the 
Pulsar Wind Nebula (PWN) Vela-X, 
described in a Science paper \cite{Pellizzoni2010},  %(Pellizzoni et al., 2010), 
has been the first experimental confirmation of $\gamma$-ray emission (E$>100$ MeV) 
from a pulsar wind nebula.
This result constrains the particle population responsible for the GeV 
emission and establishes a class of $\gamma$-ray emitters that could 
account for a fraction of the unidentified galactic $\gamma$-ray sources.
Subsequently the NASA Fermi satellite has confirmed the 
Vela-X $\gamma$-ray detection \cite{Abdo2010}, 
and has also firmly identified about 9 other pulsar wind nebulae
plus few other PWN candidates \cite{3FGL}. 

\item {\bf $\eta$-Carinae}:
AGILE has provided the first detection of a colliding wind 
binary (CWB) system above 100 MeV in the $\eta$-Carinae region \cite{Tavani2009c}.
% Tavani, M., Sabatini, S., Pian, E. et al. 2009, ApJ, 698, L142
AGILE detected a $\gamma$-ray source (1AGL J1043-5931, now 1AGLR J1044-5944)
consistent with the position of the CWB massive system $\eta$-Car during the time period
2007 July - 2009 January.
A 2-day $\gamma$-ray flaring episode was also reported on 2008 Oct. 11-13, 
possibly related to a transient acceleration 
and radiation episode of the strongly variable shock in the system.
A revised AGILE $\gamma$-ray source list in the complex Carina region 
has been published in \refcite{Verrecchia2013}. 

\item {\bf Bright $\gamma$-ray flaring blazars}: 
As it has been observed by EGRET and confirmed by Fermi, AGILE
detects only few objects with flux 
greater than $100 \times 10^{-8} {\rm ~ph~cm^{-2}~s^{-1}}$ above 100 MeV. 
Whether this is due to selection effects or there is a subclass of
blazars with peculiar characteristics is still an open question.
Moreover AGILE observations have brought to light a more complex
behaviour of blazars with respect to the standard models, indicating
the presence of two emission components in any BL Lacs,
and the possible contributions of an hot corona as source of seed
photons for the External Compton in FSRQs. 
The study of multiwavelength correlations
is the key to understand the structure of the inner jet and the origin of the
seed photons for the Inverse Compton process. 

AGILE observations of $\gamma$-ray flares from FSRQs such as 
3C 454.3 (see \refcite{Vercellone2011} and references therein), %Vercellone, S., Striani, E., Vittorini, V., et al. 2011, ApJ, 736, L38
3C 279 (see \refcite{Pittori2018,Giuliani2009} and references therein), %Pittori C. et al., 2018. to appear in ApJ.
%Giuliani, A. et al., 2009, A&A, 494, 509
PKS 1830-211(see \refcite{Donnarumma2011} and references therein), %Donnarumma, I. et al., 2011, ApJL, 736, L30
4C +21.35 (see \refcite{Verrecchia2014} and previous AGILE Astronomer’s Telegrams),
%Verrecchia, F., Lucarelli, F., Pittori, C., et al. 2014, The Astronomer’s Telegram, 6733
often extending up to TeV energies, with fast timescale variability of the order of hours or 
even minutes, together with their multi-wavelength behavior showing in general 
a very high Compton dominance, challenge simple one-zone leptonic theoretical models
(see \refcite{Tavani2015} and references therein).  %Tavani, M., Vittorini, V., & Cavaliere, A. 2015, ApJ, 814, 51

\item {\bf MCW 656:} 
AGILE detected the transient source AGL J2241+4454 in 2010,
which triggered the study of the AGILE position error box field, and led to the subsequent discovery of the
first  ``hidden black hole''  MCW 656 in a Be star binary \cite{Munar2016}.
% Munar-Adrover  P.,  et al. 2016, ApJ, 829, 101

\item {\bf Terrestrial Gamma-Ray Flashes}: \\
Surprises also came from the Earth atmosphere.
The AGILE Minicalorimeter is also 
detecting Terrestrial Gamma-Ray Flashes (TGFs),
intense and brief pulses of $\gamma$-rays 
originating in the Earth atmosphere, and associated with thunderstorm activity.
TGFs last a 
few thousandths of a second, and produce 
$\gamma$-ray flashes up to 100 MeV, on timescales as low as
$< 5$ ms \cite{Marisaldi2010}.  %(Marisaldi et al., 2010a).
AGILE data have shown for the first time that TGF cumulative spectrum at
high energy deviates from a power law with exponential cutoff model and can be better
fit with a broken power law with significant counts above background up to 100 MeV.
The crucial AGILE contribution to TGF science is thus the discovery
that the TGF spectrum extends well above 40 MeV, and that
the high energy tail of the TGF spectrum is harder than expected. 

\end{itemlist}

\subsection{AGILE and gravitational waves}
\label{sec:gw}
%The short GRB 090510 and the GW counterpart hunt. \\

\begin{figure}[pt]
%\vspace*{-48pt}
\centerline{\psfig{file=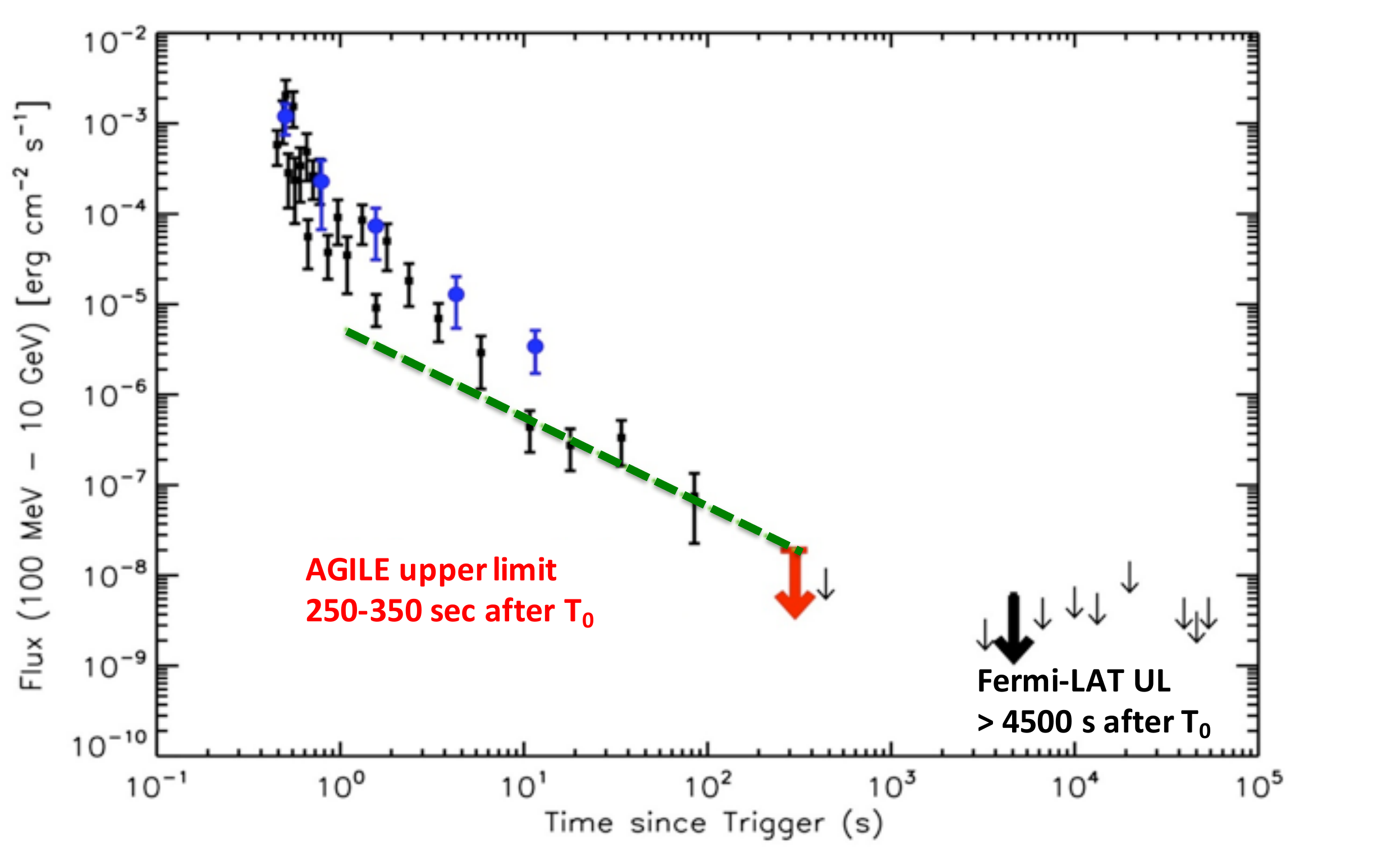,width=12cm}}
%\vspace*{-48pt}
\caption{The AGILE (blue circles) and Fermi-LAT (black squares) gamma-ray lightcurves
of the short GRB090510 scaled in flux and time as if it
originated at the GW event GW150914 luminosity distance. The AGILE-GRID upper limit 
to $\gamma$-ray emission above 100 MeV is shown in red, the corresponding Fermi-LAT upper limit
is marked in black.
\label{figure5}
}
\end{figure}

The very fast AGILE ground segment alert system \cite{Pittori2013,Bulgarelli2014}
% Pittori, C. 2013, Nuclear Physics B Proceedings Supplements, 239, 104
% Bulgarelli, A., Trifoglio, M., Gianotti, F., et al. 2014, ApJ, 781, 19
has been recently further optimized for the search of 
electromagnetic counterparts of gravitational waves, 
allowing the AGILE Team to perform a full data reduction and the preliminary Quick Look
scientific analysis only 25/30 minutes after the telemetry download from the spacecraft.
The short GRB 090510 \cite{Giuliani2010,Ackermann2010} 
%Giuliani, A., Fuschino, F., Vianello, G., et al. 2010, ApJ, 708, L84 
%Ackermann et al. 2010.
has been considered a reference for potential electromagnetic $\gamma$-ray emission 
that could be associated
to a GW event, and its lightcurve has been used as a possible high-energy template counterpart of GW events
\cite{Ackermann2016,Tavani2016}, % Ackermann, M., et al. 2016, ApJ, 823, L2
In Fig. \ref{figure5} we show in red the prompt AGILE upper limit in the case of GW150914
compared to a rescaled gamma-ray light curve of GRB 090510 (originally at z = 0.9, scaled in flux and time 
corrected as if it originated at the GW150914 luminosity distance z = 0.09).
The AGILE observations have been significant in providing the fastest response 
above 100 MeV to GW150914 and all to other GW events detected up to now with optimal gamma-ray sensitivity
\cite{Tavani2016,Verrecchia2017a,Verrecchia2017b}.

The prospects for future follow-up gamma-ray observations of GW sources by AGILE are very promising.

\section*{Acknowledgments}
The author would like to acknowledge the financial support of ASI under contract to INAF: ASI 2014-049-R.0 dedicated to SSDC.

%\begin{thebibliography}{000} %for 3 digits
%\begin{thebibliography}{00}  %for 2 digits

\end{document}